\documentclass[10pt,letterpaper]{article}
\usepackage[usenames,dvipsnames]{color}
\usepackage{opex3}
\usepackage{amsmath}
\usepackage{upgreek}
\usepackage{epstopdf}

\begin{document}

\title{Optical Eigenmodes ; \\
Exploiting the quadratic nature of the energy flux and of scattering interactions}
\author{M. Mazilu, J. Baumgartl, S. Kosmeier, and K. Dholakia}
\address{SUPA, School of Physics and Astronomy, University of St Andrews, North Haugh, Fife, KY16~9SS, UK}

\email{michael.mazilu@st-andrews.ac.uk} 

\begin{abstract}
We report a mathematically rigorous technique which facilitates the optimization of various optical  properties of electromagnetic fields in free space and including scattering interactions. The technique exploits the linearity of electromagnetic fields along with the quadratic nature of the intensity to define specific Optical Eigenmodes (OEi) that are pertinent to the interaction considered. Key applications include the optimization of the size of a focused spot, the transmission through sub-wavelength apertures, and of the optical force acting on microparticles. We verify experimentally the OEi approach by minimising the size of a focused optical field using a superposition of Bessel beams.
\end{abstract}


\section{Introduction}

The decomposition of fields into eigenmodes is a well established technique to solve various problems within physical sciences. The most prominent example is the Schr\"{o}dinger's equation within the field of quantum mechanics, where energy spectra of atoms are determined via the eigenvalue spectra and associated wavefunctions of the Hamiltonian operator. Indeed, electron orbits are eigenmodes of the energy, angular momentum, and spin operators~\cite{cohen-tannoudji1977} and as such they deliver fundamental insights into the physics of atoms. Within classical mechanics, modes of vibration of music instruments give, for example, their resonant frequencies while their spectrum is associated with the shape of the instrument~\cite{Kac:1966p9889}. In the optical domain, mode decomposition is used in order to describe light propagation within waveguides~\cite{Sudbo:1993p9727}, photonic crystals~\cite{Bienstman:2001p9890}, optical cavities~\cite{Reithmaier:1997p9888}, laser resonators~\cite{Kogelnik:1966p9295}, and the optical forces on Mie-sized particles~\cite{Barton:1989p1732}. In the case of waveguides and photonic crystals, for example, eigenmodes describe electromagnetic fields that are invariant in their intensity profile as they propagate along the fibre or crystal. Additionally, these modes are orthogonal and as such light coupled to one of these modes remains, in theory, in this mode forever.  This optical mode decomposition can be expanded to include additional properties such as orbital and spin angular momentum~\cite{Mazilu:2009p9267}.  All-together, the eigenmode expansion method is a well-established method for the representation of the propagation of optical fields.

In this paper, we report a novel method which we term ``Optical Eigenmodes (OEi)'' which represents a generalization of the powerful concept of eigenmode decomposition  going beyond the propagation properties of light. Crucially, we show that eigenmode decomposition is applicable to the case of any quadratic measure which is defined as a function of the electromagnetic field. Prominent examples of optical quadratic measures include the energy density and the energy flux of electromagnetic fields. The OEi method makes it possible to describe an optical system and its response to incident electromagnetic fields as a simple mode coupling problem and to determine the optimal ``excitation'' for the given measure considered. Intuitively, a superposition of initial fields is optimized in a manner that the minimum/maximum measure is achieved. For instance, the transmission through a pinhole is optimized by maximizing the energy flux through the pinhole~\cite{GarciaVidal:2005p9724}. 

From a theoretical perspective, the OEi optimization method is mathematically rigorous and may be distinguished  from the multiple techniques currently employed ranging from genetic algorithms~\cite{Assion:1998p9709} and random search methods~\cite{Dennis:2010p9723} to direct search methods~\cite{Thomson:2008p8158}. The major challenge encountered in any such approximate optimization and engineering of optical properties is the fact that electromagnetic waves interfere. As such the interference pattern not only makes the search for an optimum beam problematic but crucially renders the superposition found unreliable, as the different algorithms may converge on different local minima which are unstable with respect to the different initial parameters of the problem. In contrast, our proposed OEi method yields a unique solution to the problem and directly determines the optimum (maximal/minimal) measure possible.

In the first part of the paper, we introduce the OEi method and show its properties in a general context of optimizing the quadratic measures of interfering waves. In the second part, we apply the OEi formalism to minimize the focal spot size and discuss the appearance of superoscillating fields. For these applications, we describe, respectively, the electromagnetic field as a superposition of scalar Laguerre-Gaussian beams, vectorial Bessel beams or more general plane waves within the angular spectral decomposition representation of light. In the third part of the paper, we report a particular experimental implementation of the OEi method using computer controlled spatial light modulators to squeeze the spot size of a superposition of Bessel beams. In the last part of the paper, the method is applied in numerical 3D modelling to determine the OEi yielding the largest transmission through a sub-wavelength aperture and the largest optical force on a micrometer sized particle. The paper concludes with a discussion of the particular results obtained and with general comments on the versatility of the OEi method to a wide range of problems. A short annex compares the convergence properties of the OEi approach with standard phase front correction methods.


\section{Method} 
\label{s:fundcon}

The OEi method exploits both the linearity of Maxwell's equations and the quadratic dependence of light-matter interactions on the electromagnetic field $\{\mathbf{E},\mathbf{H}\}$ where $\mathbf{E}$ and $\mathbf{H}$ denote the electric and magnetic field vectors, respectively. Table~\ref{tab} provides a list of common examples of such interactions. These interactions may be written in a general quadratic matrix form
\begin{equation}
\label{eq:OEi1}
{m}^{(A)}(\mathbf{E},\mathbf{H})=\mathbf{a}^{\dagger}\mathbf{M}^{(A)}\mathbf{a}
\end{equation}
where we considered a superposition of fields $\left\{\mathbf{E},\mathbf{H}\right\}  =\left\{\sum_{j=1}^{N}a_{j}\mathbf{E}_{j},\sum_{j=1}^{N}a_{j}\mathbf{H}_{j}\right\}$ and where $(A)$ labels the light-matter interaction defined in Table~\ref{tab}. The vectors $\mathbf{a}$ and $\mathbf{a}^{\dagger}$ are  comprised of the superposition coefficients $a_{j}$ and their complex conjugates, respectively. The elements $M_{jk}^{(A)}$ of $\mathbf{M}^{(A)}$ are constructed by combining the respective fields $\{\mathbf{E}_{j},\mathbf{H}_{j}\}$ and $\{\mathbf{E}_{k},\mathbf{H}_{k}\}$ for $j,k=1\dots N$. 
More precisely, we have:
\begin{eqnarray}
\label{eq:matrix}
4M_{jk}^{(A)}&=&{m}^{(A)}(\mathbf{E}_j+\mathbf{E}_k,\mathbf{H}_k+\mathbf{H}_k)-i{m}^{(A)}(\mathbf{E}_j+i\mathbf{E}_k,\mathbf{H}_k+i\mathbf{H}_k) \nonumber \\
&&-{m}^{(A)}(\mathbf{E}_j-\mathbf{E}_k,\mathbf{H}_k-\mathbf{H}_k)+i{m}^{(A)}(\mathbf{E}_j-i\mathbf{E}_k,\mathbf{H}_k-i\mathbf{H}_k). 
\end{eqnarray}
Given the Hermitian form of~\eqref{eq:matrix}, we remark that the light-matter interaction $\mathbf{M}^{(A)}$ defines a spectrum of real eigenvalues $\lambda_{k}^{(A)}$ and associated eigenvectors $\mathbf{v}_{k}^{(A)}$. Each of these eigenvectors corresponds to a superposition of fields 
$\{\mathbf{E}_{j},\mathbf{H}_{j}\}$ termed here Optical Eigenmode (OEi). Crucially, we may now extremize the light-matter interaction considered; that is, the extremal eigenvalue $\lambda_{\textrm{ext}}^{(A)}$ and the associated eigenvectors $\mathbf{v}_{\textrm{ext}}^{(A)}$ deliver the superposition of fields $\left\{\mathbf{E}_{\textrm{ext}},\mathbf{H}_{\textrm{ext}}\right\}=
\left\{\sum_{j=1}^{N}v_{\textrm{ext},j}^{(A)}\mathbf{E}_{j},\sum_{j=1}^{N}v_{\textrm{ext},j}^{(A)}\mathbf{H}_{j}\right\}$ which extremizes the interaction $m^{(A)}$ while keeping the total field contributions constant $\mathbf{a}^{\dagger}\mathbf{a}=1$.

In this paper, we apply the OEi concept to minimize the size of a laser spot within a surface ROI. One way to define the spot size of a laser beam is by measuring (whilst keeping the total intensity constant) the second order moment $w$ of its intensity  distribution~\cite{Paschotta2008EncOfLasPhysAndTechn}. Crucially, $w$ can be expressed in terms of $m^{(0)}$ and $m^{(2)}$ (see Table~\ref{tab}) or the respective matrix representations \eqref{eq:OEi1} as follows: 
\begin{equation}
\label{eq:OEi2}
w = 2\sqrt{\frac{m^{(2)}}{m^{(0)}}}=
2\sqrt{\frac{\mathbf{a}^{\dagger}\mathbf{M}^{(2)}\mathbf{a}}{
\mathbf{b}^{\dagger}\mathbf{M}^{(0)}\mathbf{b}}},
\end{equation}
where $\mathbf{M}^{(0)}$ and $\mathbf{M}^{(2)}$ are termed the \emph{intensity operator (IO)} and \emph{spot size operator (SSO)}, respectively. According to~\eqref{eq:OEi2}, the minimum spot size is obtained by the OEi associated with the smallest eigenvalue $\lambda^{(2)}$ of the SSO provided that the IO is simultaneously diagonalized and normalized to 1. Direct evaluation shows that this is precisely achieved by the combined OEi 
\begin{equation}
\label{eq:OEi3}
\left\{\mathbf{E}_{\min},\mathbf{H}_{\min}\right\}
=\left\{\sum_{j=1}^{N}\sum_{k=1}^{N}\frac{v_{\min,k}^{(2)}v_{k,j}^{(0)}}{\sqrt{\lambda_{k}^{(0)}}}\cdot\mathbf{E}_{j}, 
\sum_{j=1}^{N}\sum_{k=1}^{N}\frac{v_{\min,k}^{(2)}v_{k,j}^{(0)}}{\sqrt{\lambda_{k}^{(0)}}}\cdot\mathbf{H}_{j}\right\}.
\end{equation} 
where $v_{\min,k}^{(2)}$ is the eigenvector associated with the smallest eigenvalue of $\mathbf{M}^{(2)}$ and in the intensity normalised eigenbase $v_{k,j}^{(0)}/{\sqrt{\lambda_{k}^{(0)}}}$ of  $\mathbf{M}^{(0)}$. 

\begin{table}[htb]
\begin{center}
\begin{tabular}{|r|l|}
\hline
Energy &  $m^{({\cal E})}(\mathbf{E},\mathbf{H})=\frac{1}{2}\int_{V}{\cal E}\;dv$ \\
Intensity &  $m^{(\textrm{0})}(\mathbf{E},\mathbf{H})=\frac{1}{4} \int_S\left( \mathbf{E}^*\times\mathbf{H}\right)\cdot\mathbf{n}\;d\sigma$+c.c. \\
Spot size  &   $m^{(\textrm{2})}(\mathbf{E},\mathbf{H})=\frac{1}{4}\int_S\mathbf r^2 \left( \mathbf{E}^*\times\mathbf{H}\right) \cdot \mathbf{n}\;d \sigma $+c.c. \\
Momentum  & $m^{(\mathbf{F\cdot u})}(\mathbf{E},\mathbf{H})= \frac{1}{4}\int_S (\epsilon_0 ({\bf E}^*\cdot {\bf n}) {\bf E}+\mu_0({\bf H}^*\cdot{\bf n}){ \bf H}-\frac{1}{2}{\cal E} {\bf n})\cdot {\bf u} \;d \sigma $+c.c. \\[1mm]
\hline
\end{tabular}
\caption{Time averaged quadratic measures $m$ of common light-matter interactions. The integration either over a volume $V$ or a surface $S$ which in general corresponds to the $\textrm{Range of interest}=\textrm{ROI}$ of the measure.  In the optical momentum case, it corresponds to a closed surface surrounding the scattering object with $\bf F \cdot \bf u$ representing the optical force in the direction defined by the unit vector $\bf u$. For surface integrals, $\bf n$ is the normal unit vector to the surface considered. The definition of the electromagnetic energy density is ${\cal E}=1/2 (\epsilon_0\bf E\cdot \bf E^*+\mu_0\bf H\cdot \bf H^*)$. }
\label{tab}
\end{center}
\end{table}

For our proof-of-principle studies described in the remainder of the paper, we also applied the scalar version of the OEi method where a set of scalar fields $E_{i}$ is considered in order to determine the IO and SSO as 
\begin{equation}
\label{eq:OEi4}
M_{jk}^{(0)}=\int_S E_{j}^{\ast}E_{k}\;d\sigma
\end{equation}
and
\begin{equation}
\label{eq:OEi5}
M_{jk}^{(2)}=\int_S \mathbf{r}^{2}E_{j}^{\ast}E_{k}\;d\sigma,
\end{equation}
respectively. These scalar expressions are equivalent to the respective vector versions listed in Table~\ref{tab} and determined through equation~\eqref{eq:matrix}. The scalar version of the optimimum OEi~\eqref{eq:OEi4} explicitly reads
\begin{equation}
\label{eq:OEi6}
E_{\min}=\sum_{j=1}^{N}\sum_{k=1}^{N}\frac{v_{\min,k}^{(2)}v_{k,j}^{(0)}}{\sqrt{\lambda_{k}^{(0)}}}\cdot E_{j}.
\end{equation}

\subsection{Smallest focal spot using Laguerre Gaussian beams}
\label{s:sec33}

\begin{figure}[htb]
\centering\includegraphics[width=12cm]{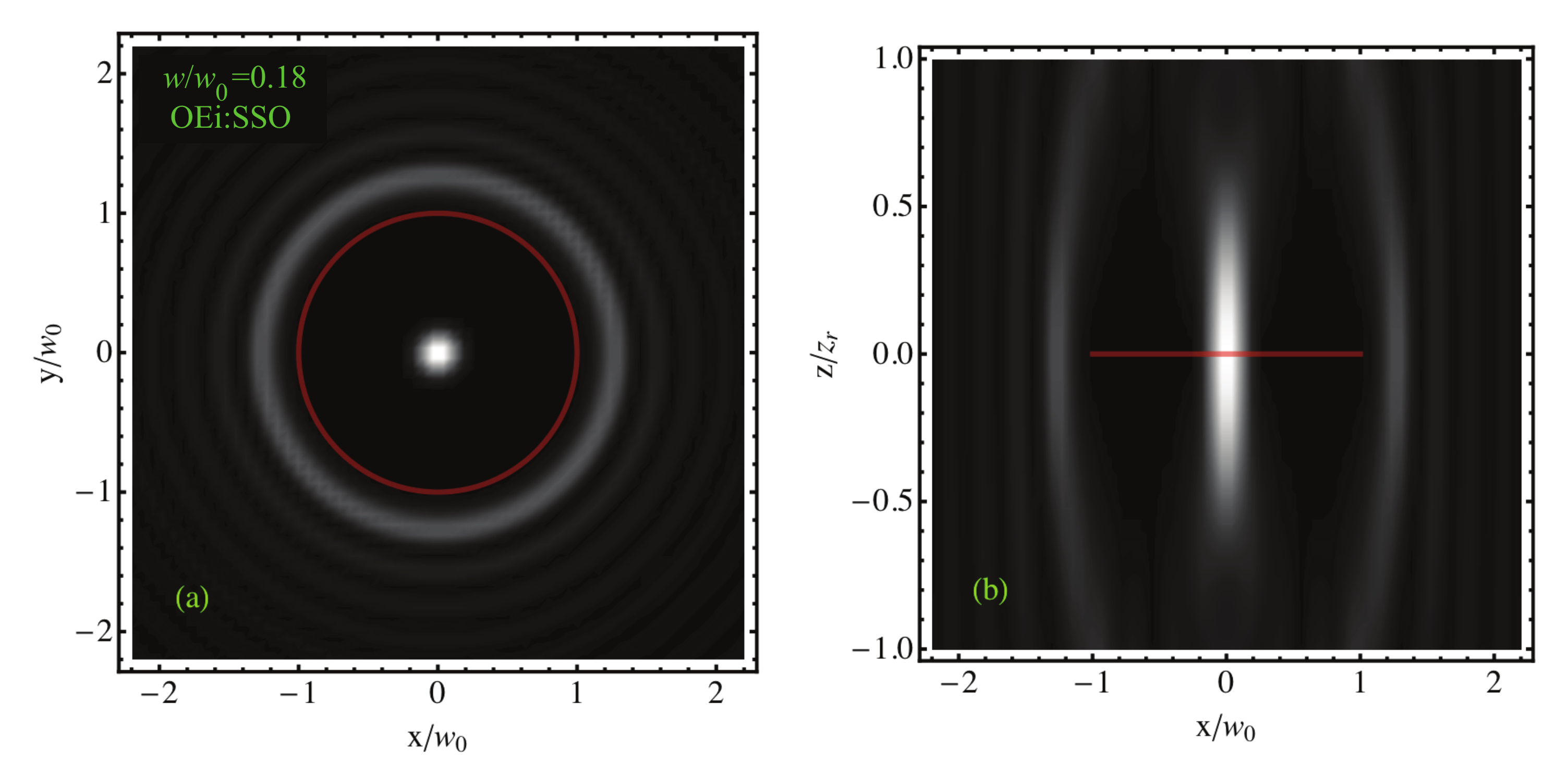}
\caption{(a) Transversal and (b) longitudinal 2D intensity cross sections of the OEi superposition delivering the smallest focal spot in the ROI ($R=\lambda$) considering 25 LG modes. $w/w_0$ is the relative spot size measured according to Eq.~\eqref{eq:OEi2}. The Strehl ratio in (a) is $4.5\%$.}
\label{fig3}
\end{figure}

\begin{figure}[htb]
\centering\includegraphics[width=13cm]{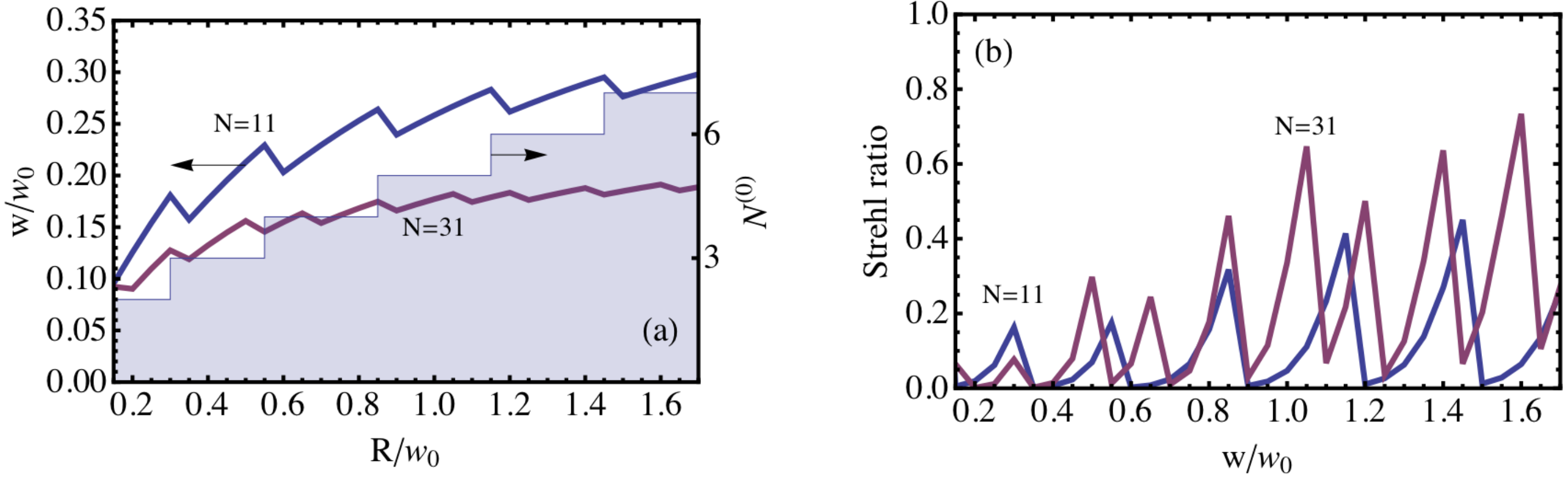}
\caption{(a) Spot size as a function of the radius of the ROI for different number of LG modes considered. The right hand scale and filled curve indicate the numbers of intensity eigenmodes $N^{(0)}$ fulfilling the intensity criteria for the $N=11$ case. The arrows indicate the corresponding scales. (b) Ratio between the ROI intensity of the smallest spot size eigenmode and the largest intensity achievable in the ROI (Strehl ratio).}
\label{fig4}
\end{figure}

Using a superposition of LG beams we can minimize the size of a focal spot using the representation of the SSO \eqref{eq:OEi5} in cylindrical coordinates. It is important to note at this point that we only retain the intensity OEis whose eigenvalues are within a chosen fraction of total intensity. This is equivalent to considering only beams that have a significant intensity contribution in the ROI. Intuitively, the optimization procedure may perform so well that a spot of size zero is finally obtained if no intensity threshold is applied. Figure~\ref{fig3} shows the smallest spot superposition where we observe the appearance of sidebands just outside the ROI.  These sidebands are a secondary effect of squeezing the light below its diffraction limit. It is these sidebands that decrease the efficiency of the squeezed spot with respect to the maximal possible intensity in the ROI as calculated via the IO. Using the ratio between these two intensities we can define the intensity Strehl ratio~\cite{Sales:1997p9891} for the SSO (see Fig.~\ref{fig4}b). We remark that both, the spot size and the Strehl ratio, show resonances as a function of the ROI size. This can be explained by considering the number of intensity eigenmodes used for the spot size operator. Indeed, as the ROI size decreases, so does the number of significant intensity eigenmodes. Each time one of these modes disappears (step in Fig.~\ref{fig4}), we have a sudden increase in the minimum spot size achievable accompanied with an enhanced Strehl ratio as we drop the most intensity inefficient mode. Overall, the Strehl ratios determined in our studies predominantly exceeded values of $1\%$ even when spots were tightly squeezed. Therefore, the observed decrease of intensity is not to severe in terms of potential applications of squeezed beams such as optical manipulation and imaging. 

\begin{figure}[htb]
\centering\includegraphics[width=12cm]{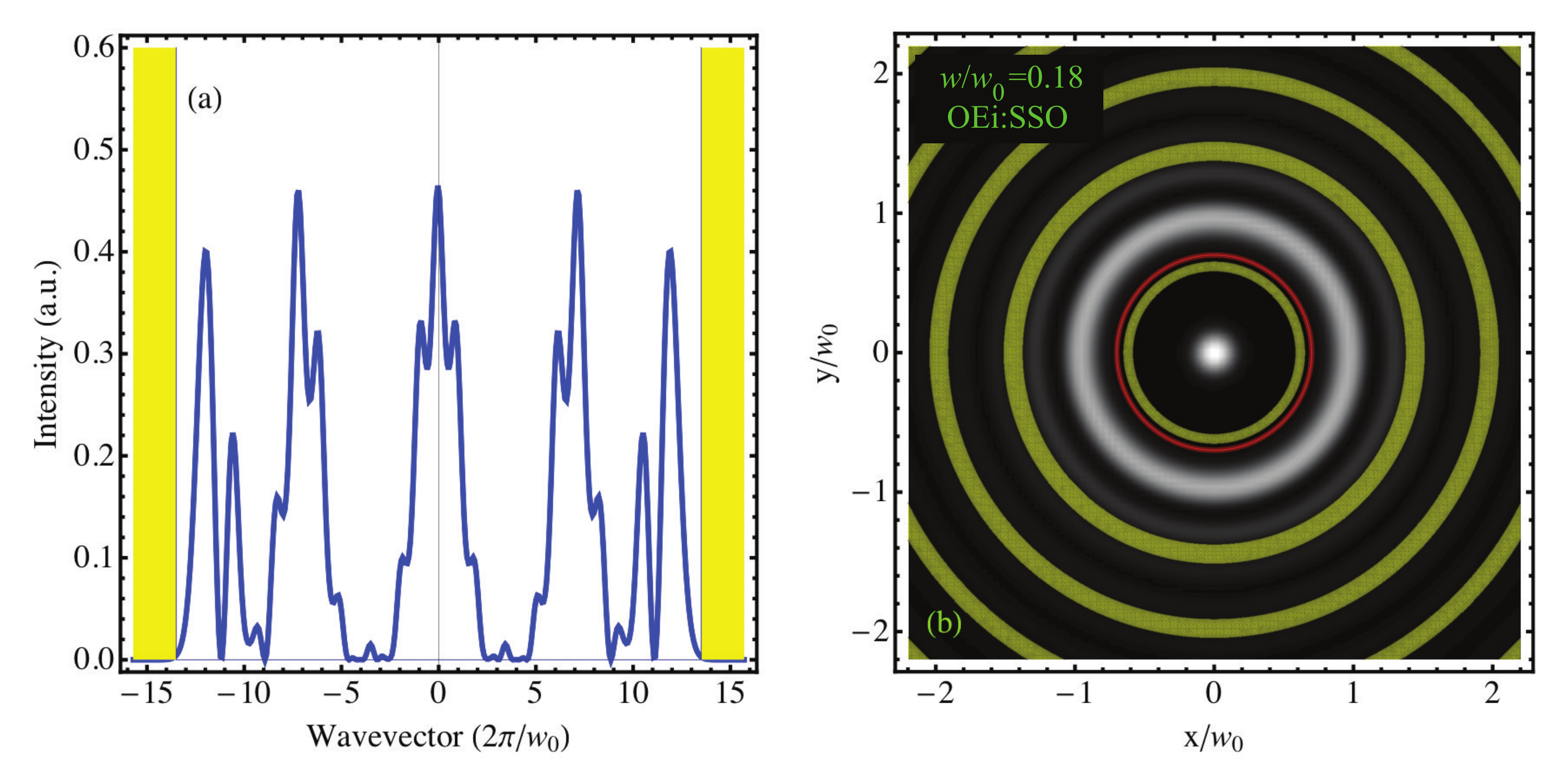}
\caption{(a) Radial wavevector spectral density. Yellow highlights regions outside the spectral bandwidth. (b) Transversal cross section of the OEi spot size optimized field intensity with yellow showing super-oscillating regions.}
\label{fig6}
\end{figure}

On a final note, we remark that squeezing light below its diffraction limit may be associated with the effect of super-oscillations~\cite{Berry:2006p2578}. This refers specifically to the ability to have a local $k$-vector (gradient of the phase) larger than the spectral bandwidth of the original field. To visualize this effect, in the case of OEi spot size optimized beams, we have calculated the spectral density of the radial wave-vector for the smallest planar spot~\cite{Dennis:12p9892}. As shown in Fig.~\ref{fig6}, this spectral density clearly identifies a spectral bandwidth (white background in Fig.~\ref{fig6}). Regions of the beam which exhibit locally larger wave-vectors than the ones supported by this spectral band width correspond to super-oscillating regions. The local wave vector is defined as $\partial_r \arg(u(r))$ where $\arg(u)$ defines the phase of the analytical signal $u$. In this particular case, we observe that super-oscillations occur in the dark region of the beam. Additionally, when the ROI is large compared to the Gaussian beam waist $w_0$, there are no super-oscillating regions. These only appear when the beam starts to be squeezed.

\subsection{Smallest focal spot using Bessel beams}
\label{s:sec34}

\begin{figure}[htb]
\centering\includegraphics[width=11cm]{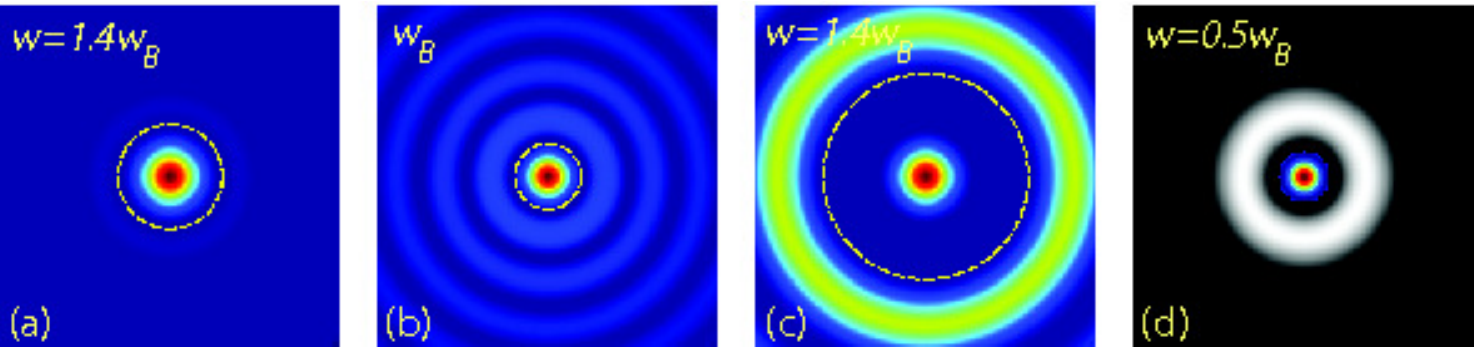}
\caption{Intensity cross sections: (a) Airy disk for the maximum numerical aperture considered NA$=\sin(\theta_{max})=0.1$. The yellow dashed circle shows the position of the smallest zero-intensity circle taken as the ROI inside which the spot size is calculated. The spot size is normalized to the spot size of the reference Bessel beam. (b) Reference Bessel beam corresponding to the largest cone angle $\theta_{max}$. The spot size of the reference Bessel beam is denoted as $w_{\textrm{B}}$. (c) OEi spot size optimized beam for a superposition of Bessel beams ($\theta\in[0,\theta_{max}]$) for a large ROI highlighted by the dashed yellow circle. Strehl ratio: $2\%$. (d) OEi spot size optimized beam for a small ROI. Strehl ratio: 0.$2\%$. The gray-scaled region shows the sidebands while the color range the ROI. Notice that the two scales are different.}
\label{fig7}
\end{figure}

\begin{figure}[htb]
\centering\includegraphics[width=12cm]{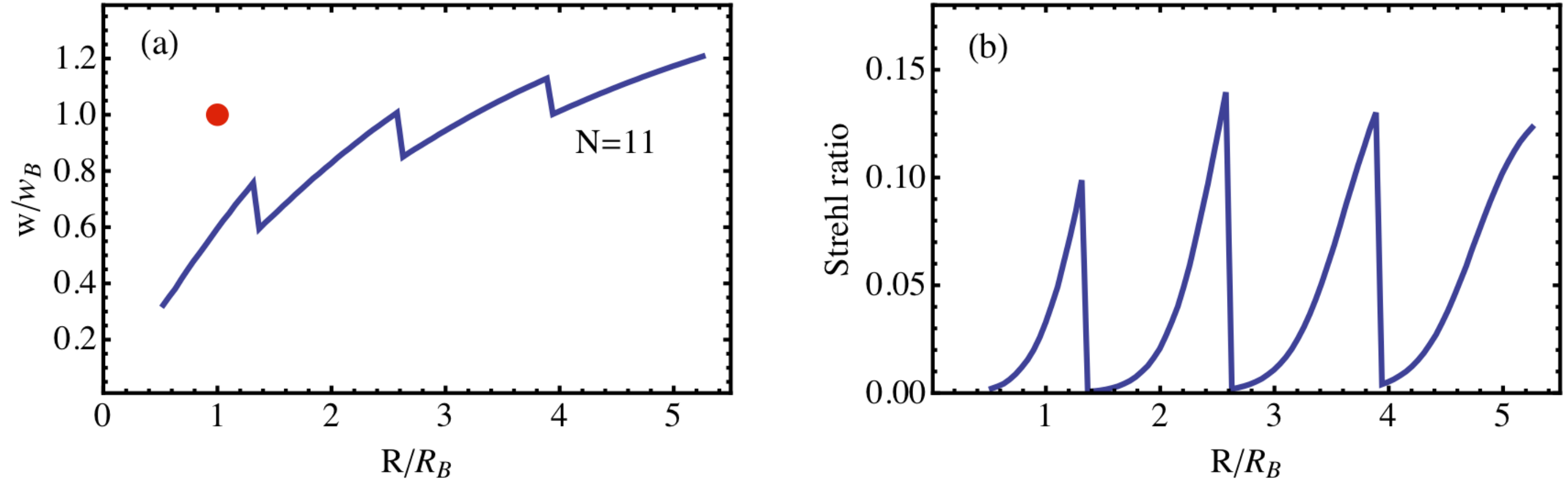}
\caption{(a) Relative spot size $\Delta r/w_{\textrm{B}}$ of the Bessel beam superposition as a function of the relative ROI radius $R/R_{\textrm{B}}$. The spot size $w_{\textrm{B}}$ and the ROI radius $R_{\textrm{B}}$ are associated with the reference Bessel beam shown in Fig.~\ref{fig7}(b), where the ROI is indicated as dashed circle. For comparison, the red dot indicates the location of the reference beam in the $\Delta r/w_{\textrm{B}}$ vs. $R/R_{\textrm{B}}$ plot. (b) Strehl ratio vs relative ROI radius $R/R_{\textrm{B}}$.}
\label{fig8}
\end{figure}

The paraxial approximation employed above in the case of LG beams can be used to describe sub-diffracting beams but breaks down when beams are tightly focused. As a consequence we must consider full vectorial solutions of Maxwell's equations. Here, we have chosen Bessel beams as a basis-set and determined the superposition of Bessel beams which minimized the spot size in a planar finite ROI. Note that the problem of the finite intensity of Bessel beams~\cite{Durnin:1987p7503} is easily circumvented here due to the finite ROI size considered. The monochromatic electric vector field of the vectorial Bessel beam may explicitly be expressed as~\cite{Volke-Sepulveda:2002hk}
\begin{eqnarray}
\label{eq:scal-BB}
\nonumber
\mathbf{E}& =&E_0  \exp\left(i\ell\phi+ik_t z)\right) \Bigg((\alpha \mathbf{{e}}_x+\beta \mathbf{{e}}_y) J_\ell(k_t r) \\
\label{eq:s341}
&&+\frac{ik_t}{2k_z}((\alpha+i\beta)\exp(-i\phi)J_{\ell-1}(k_t r)-(\alpha-i\beta)\exp(i\phi)J_{\ell+1}(k_t r))\mathbf{{e}}_z\Bigg)
\end{eqnarray}
where $k_t=k_0\sin(\theta) $ and $k_z=k_0\cos(\theta) $ are the transversal and longitudinal wave vectors with $\theta$ the characteristic cone angle of the Bessel beam. $\mathbf{{e}}_{x}$, $\mathbf{{e}}_{y}$ and $\mathbf{{e}}_{z}$ are the unit vectors in the Cartesian coordinate system. The parameter $\ell$ corresponds to the azimuthal topological charge of the beam while $\alpha$ and $\beta$ are associated with the polarization state of the beam. The magnetic field  $\mathbf{H}$ was deduced according to Maxwell's equations. Figure~\ref{fig7} shows a comparison between the Airy disk, the Bessel beam and the OEi optimized spot considering a numerical aperture of NA$=0.1$. As in the case of the LG beams, squeezing the focal spot  is accompanied by side bands and a loss in efficiency shown by the Strehl ratio (see Fig.~\ref{fig8}). 

\section{Experimental OEi}
\label{s:sec4}

\subsection{Experimental implementation of the OEi concept}

To perform an experimental OEi optimization we have used the following setup: A $\textrm{HeNe}$ laser beam is expanded and subsequently amplitude modulated by a spatial light modulator (SLM) display operating in conjunction with a pair of crossed polarizers. Analogously to a liquid crystal display on a computer or laptop monitor, the liquid crystal SLM display rotates the polarization of the incident light by an angle depending upon the voltage applied to the display pixels. The amplitude modulated beam is then imaged onto a second SLM display through a pair of lenses. This second SLM display along with a subsequent Fourier lens and aperture served to modulate the phase of the laser beam in the standard first order configuration~\cite{DiLeonardo2007}. The field modulations of interest were encoded as RGB images where the blue channel represented the amplitude and the green channel the phase modulation. The SLM controller extracted these information and applied the two channels to the respective panel. We have performed calibration measurements to ensure that both the amplitude and phase modulation exhibited a linear dependence on the applied 8-bit color value between 0 and 255. A CCD camera allowed us to record images of laser fields in the Fourier plane of lens 5. 

\begin{figure}[htb]
\centering\includegraphics[width=0.85\textwidth]{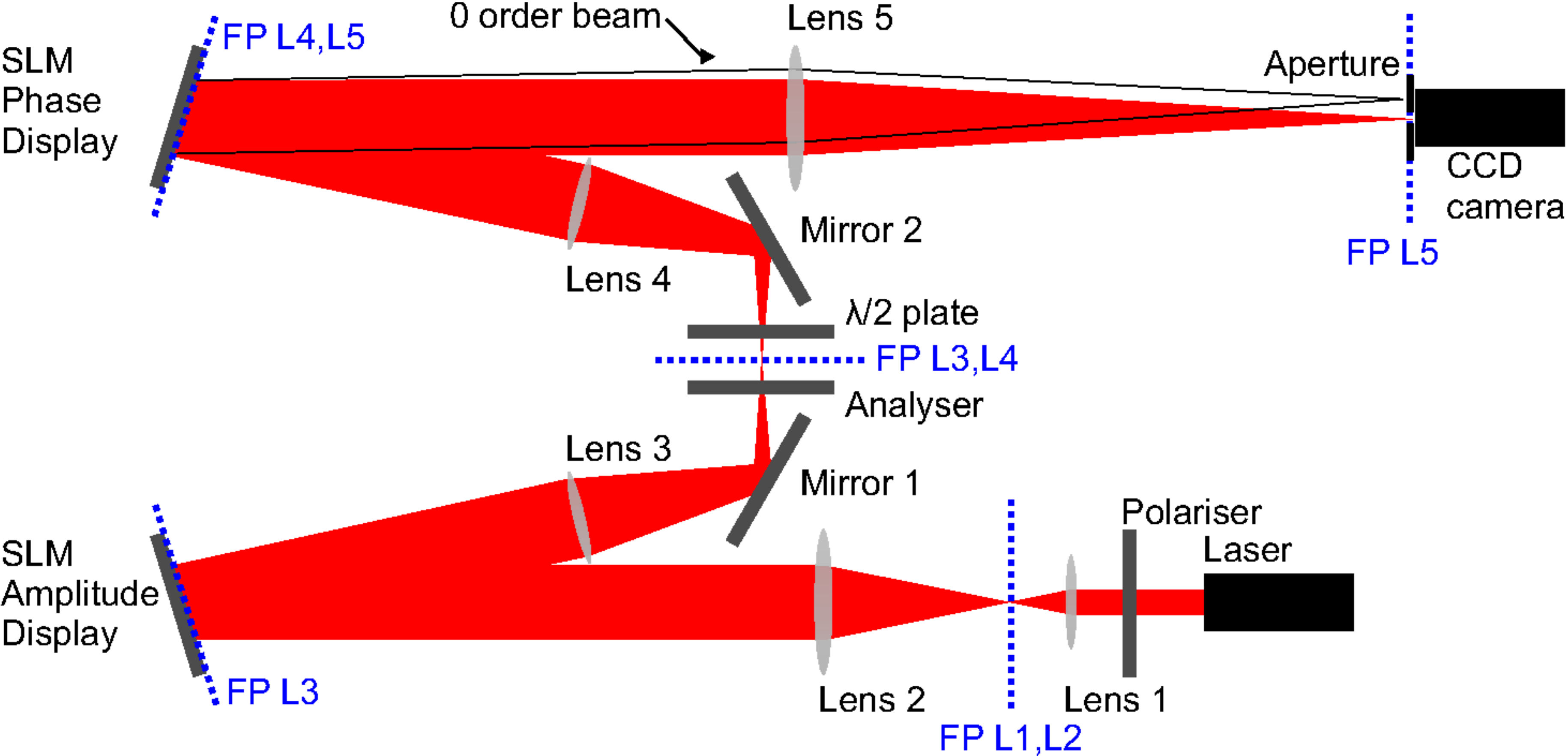}
\caption{Experimental setup. FP = focal plane, L = Lens. Focal widths: $f_{1}=50~\textrm{mm}$, $f_{2}=500~\textrm{mm}$, $f_{3}=f_{4}=400~\textrm{mm}$, $f_{5}=1~\textrm{m}$. Laser: JDS Uniphase HeNe laser, $P_{\max}=10~\textrm{mW}$, $\lambda=633~\textrm{nm}$, SLM: Holoeye HEO 1080 P dual display system, $\textrm{resolution}=1920~\textrm{pixel}\times1080~\textrm{pixel}$, $\textrm{display size}=1~\textrm{in}\times0.7~\textrm{in}$. CCD camera: Basler pilot piA640-210gm, $\textrm{resolution}=648~\textrm{pixel}\times488~\textrm{pixel}$, $\textrm{pixel size}=7.4~\upmu\textrm{m}\times7.4~\upmu\textrm{m}$.}
\label{fig:setup}
\end{figure}

The experiment consist in determining the OEi  in the CCD camera plane whilst shaping and superimposing the test fields $E_{j}$ in the SLM planes.
In the following, we indicate the plane of interest by a $z$-coordinate along the optical axis where $z=z_{1}$ and $z=z_{2}$ refer to the SLM and CCD camera plane, respectively. According to this convention we shape a set of test fields $E_{j}(z_{1})=A_{j}(z_{1})e^{i\phi_{j}(z_{1})}$ both in amplitude $A_{j}$ and phase $\phi_{j}$ in the SLM plane, and the associated intensities $I_{j}(z_{2})\propto|E_{j}(z_{2})|^{2}$ are detected in the CCD camera plane. The amplitudes $A_{j}(z_{2})$ were determined from these intensities by simply taking the square root. We used the three-step phase retrieval algorithm described in Ref.~\cite{Malacara1992OpticalShopTesting} to retrieve the phase modulations $\phi_{j}(z_{2})$. 
The determination of the phase and amplitude of the beam in the CCD plane allows us to numerically vary the ROI without redisplaying the test fields. Using these fields, the IO and SSO are determined according to Eqs.~\eqref{eq:OEi4} and~\eqref{eq:OEi5}, respectively.

During the course of our experiments we verified the linearity of our optical system by performing a comparison between what we term the ``experimental superposition (Exp-S)'' and the ``numerical superposition (Num-S)''. The Exp-S refers to the case where the set of OEi optimized  superposition coefficients $a_{i}$ is used to encode the optimized superimposed field onto the SLM. The CCD camera then detected the intensity $I_{\textrm{Exp-S}}(z_{2})$ corresponding to this encoded optimized field. The Num-S utilizes the fields $E_{j}(z_{2})$, which were individually measured to assemble the OEi operators, in order to \emph{numerically} determine the intensity distribution as $I_{\textrm{Num-S}}(z_{2})\propto\left|\sum_{i=1}^{N}a_{j}E_{j}(z_{2})\right|^{2}$. Crucially, linearity is verified if $I_{\textrm{Exp-S}}(z_{2})=I_{\textrm{Num-S}}(z_{2})$. This is indeed observed in our experiments as demonstrated in the following subsection which features a comparison of  experimental and numerical intensity distributions. 

\subsection{Results and discussion}

\begin{figure}[htb]
\centering\includegraphics[width=0.75\textwidth]{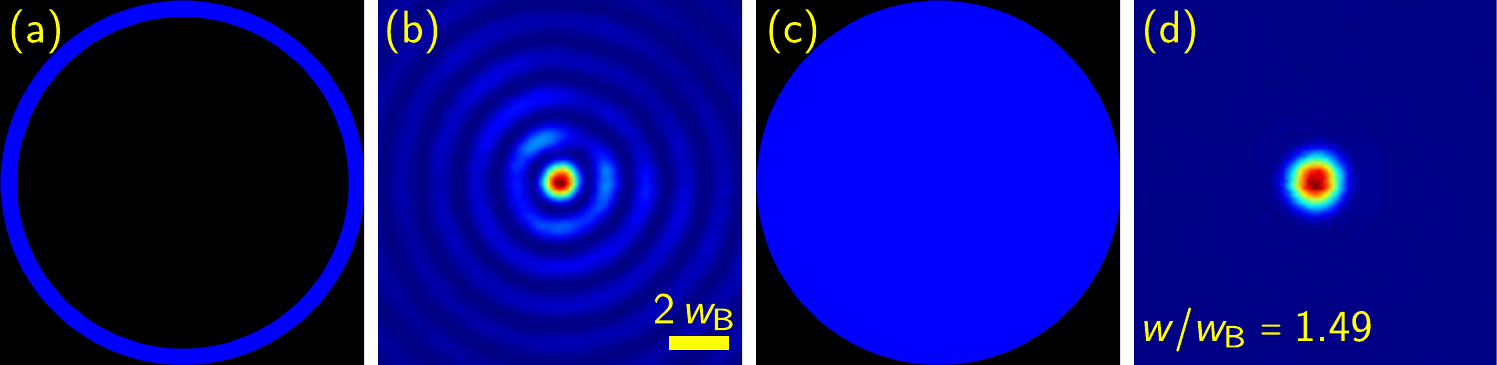}
\caption{SLM encoded field modulations and resulting beam profiles. (a) Ring mask RGB image as encoded onto the dual panel SLM. (b) Associated Bessel beam created in the CCD camera plane. (c) Aperture RGB image as encoded onto the dual panel SLM. (d) Associated Airy disk as detected by the CCD camera. The yellow bar in (b) represents 2 times the spot size $w_\text{B}$ of the Bessel beam's central core. $w$ in (d) is the spot size of the Airy disk.}
\label{fig:bessels}
\end{figure}

In our experiments, we used $N = 11$ non overlapping amplitude ring masks with a constant phase modulation as fields of interest $E_{i}(z_{1})$. After propagation through the Fourier lens 5 (see Fig.~\ref{fig:setup}) the resulting fields $E_{i}(z_{2})$ form a set of Bessel beams. Figure~\ref{fig:bessels}(a) shows the largest ring modulation encoded onto the SLM with the resulting Bessel beam shown in Fig.~(b). As this particular Bessel beam comes along with the highest NA compared to the Bessel beams created with smaller ring modulations, the beam shown in 
Fig. (b) exhibits the smallest central spot of all beams realized in our experiments. The spot size of the Bessel beam featuring the smallest core is denoted as $w_\text{B}$ and used as reference for the measurements presented below. For comparison Figure~\ref{fig:bessels}(c) depicts a circular aperture which is encoded onto the SLM in order to observe the Airy disk (see Fig.~(d)). The spot size of the Airy disk is approximately $1.5$ times larger than the core of the reference Bessel beam as expected~\cite{Wang:2008p9893}. 

\begin{figure}[htb]
\centering\includegraphics[width=0.75\textwidth]{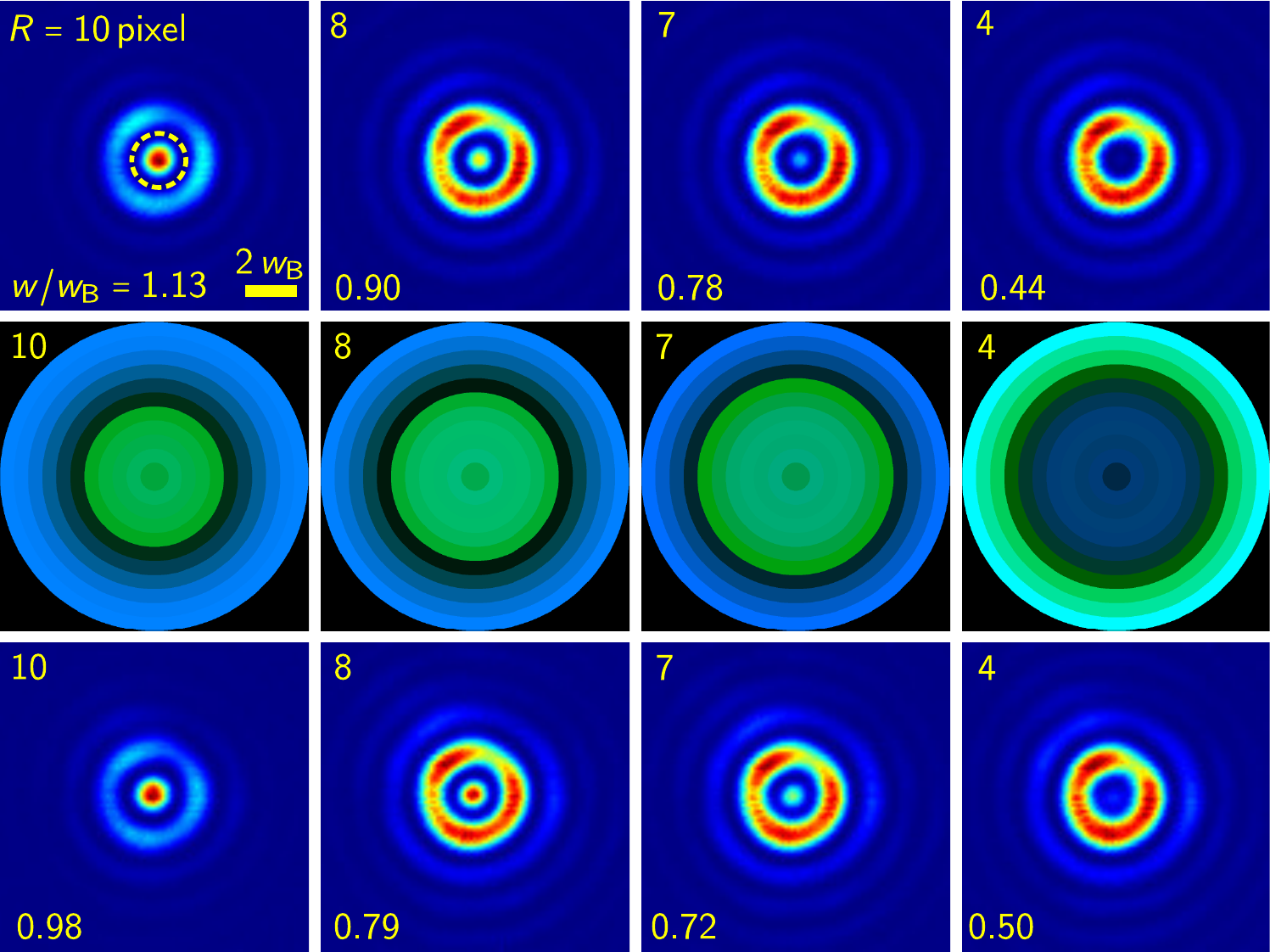}
\caption{Experimental OEi spot size minimization. \emph{Top row}: $I_{\textrm{Num-S}}(z_{2})$ for different ROI radii in pixel as indicated in the top left corner of all graphs shown. The ROI is exemplary indicated as a dashed ring in the left hand side intensity distribution. The number in the bottom left corner represents the spot size $w$ in units of the reference spot size $w_{\textrm{B}}$. \emph{Central row}: Optimized experimental distribution as RGB encoded onto the SLM. \emph{Bottom row}: Intensity distributions $I_{\textrm{Exp-S}}(z_{2})$. The relative spot size $w/w_\text{B}$ is indicated in the lower left corner.}
\label{fig:spots}
\end{figure}

The results of the performed OEi spot size minimization are shown in Fig.~\ref{fig:spots} for different sizes of the ROI. To begin with, the comparison of the Num-S intensity distribution $I_{\textrm{Num-S}}(z_{2})$ (top row) and the Exp-S intensity distributions $I_{\textrm{Exp-S}}(z_{2})$ (bottom row) clearly reveals good agreement and thus verifies the linearity of our optical system as elucidated above. For completeness, the central row shows the Exp-S superposition in RGB format as encoded onto the SLM. The color code features a blue channel representing the amplitude modulation from 0 (black) to 1 (blue) and a green channel corresponding to phase modulations from 0 (black) to $2\pi$ (green). Next, we conclude from the measured relative spot size $w/w_\text{B}$ that the spot size decreases if the ROI size is reduced. The reduced spot size is achieved at the expense of the spot intensity which is redistributed to a ring outside of the ROI similar to the theoretical results presented in section~\ref{s:sec34} and Fig.~\ref{fig7}. Referring to the Exp-S data, for $R = 7\,\text{pixel}$ the spot size is reduced to $72\,\%$ of the size of the reference Bessel beam's core and even further to $50\,\%$ for $R = 4\,\text{pixel}$. The latter result is somewhat vague, though, due to the low spot intensity which may be truncated by the sensitivity threshold of the CCD detector and thus may appear smaller. However, our experimental results overall clearly verify the OEi concept applied to spot size minimization. Moreover, the results strongly suggest that the OEi optimization may indeed squeeze spots to the subdiffractive regime since the optimal superposition of Bessel beams not only beats the Airy disk but also the reference Bessel beam diameter. 

\section{Applicability of the OEi method to scattering interactions}

\begin{figure}[htb]
\centering\includegraphics[width=10.3cm]{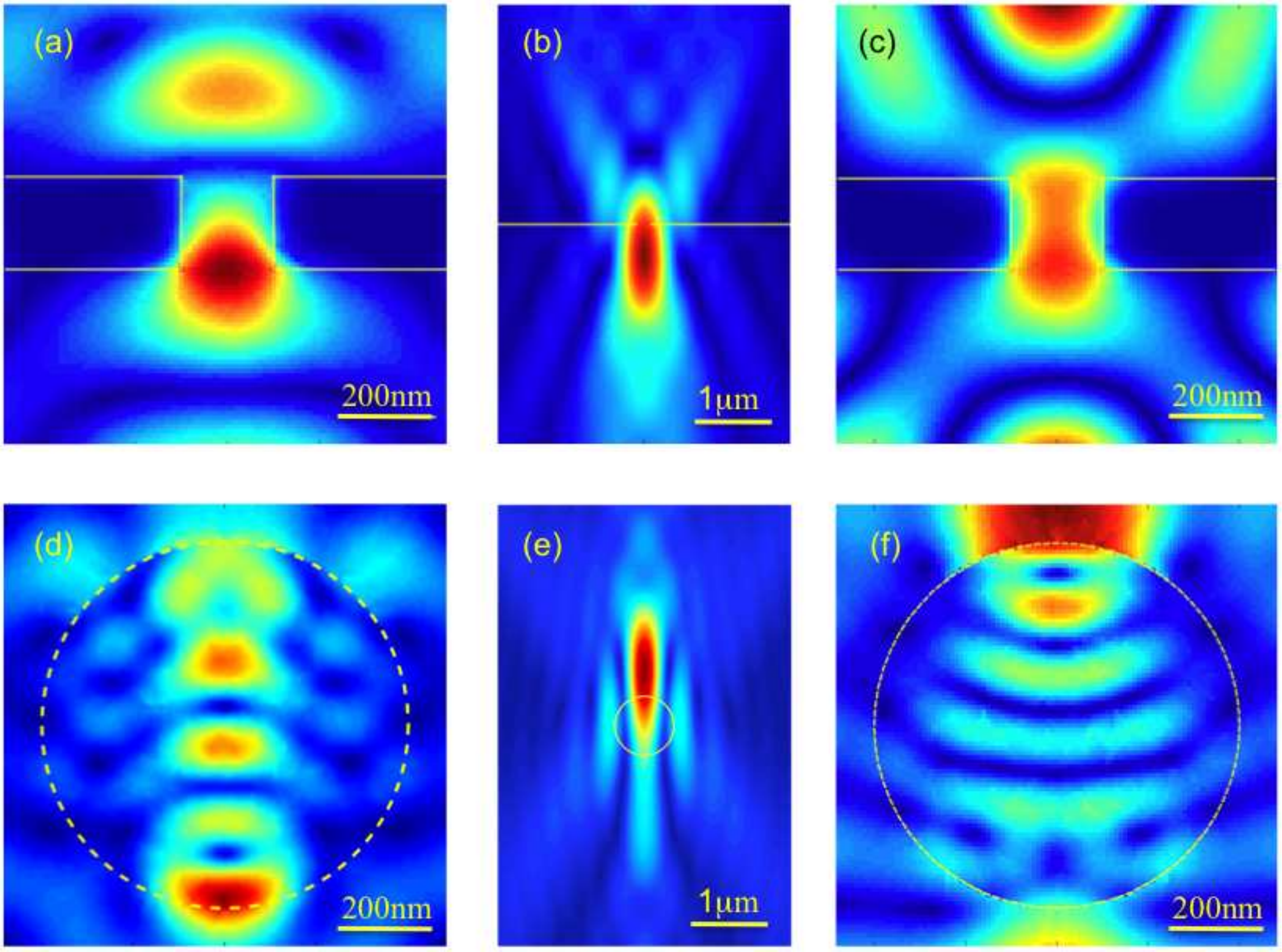}
\caption{(a-c) Cross section plot of the electric field amplitude, $|\bf E|$, for a sub-wavelength aperture (diameter=200nm) in a thin layer of silver (thickness=200nm, refractive index $n=0.12-3.7i$ at a wavelength $\lambda=600$nm) illuminated from below. The yellow lines represent the boundary of the structure. (a) Intensity OEi ensuring the largest transmission (transmission enhancement factor 2.1 with respect to the tightest Bessel beam and 1.55 with respect to the Airy disk illumination). (b) Incident intensity OEi without the structure. (c) Tightest Bessel beam illumination.  (d-f) Electric field amplitude, $|\bf E|$, in a cross section for a high refractive index ($n=1.8$) microparticle (diameter=800nm) illuminated from below with a wavelength ($\lambda=504$nm). (d) Momentum OEi ensuring the largest momentum transfer (enhancement factor 49.3 with respect to the plane wave and 1.33 with respect the Airy disk illumination). (e) Incident momentum OEi without the structure. (f) Plane wave illumination.   }
\label{fig:oei}
\end{figure}

In this section, we demonstrate, on the basis of a numerical study, how the OEi method can equally be applied in some scattering interaction processes. Indeed, the OEi method presented above is applicable to free space propagation and can be directly extended to linear scattering processes where the optical interaction is expressed as a quadratic form of the field. As in the case of the smallest spot operator, the OEi of these light matter interactions can be used to determine the electromagnetic field profile delivering the largest or the smallest interaction strength. In this section, we show two numerical examples illustrating the cases presented in Table~\ref{tab}. The numerical modelling is performed using a finite element method (Comsol) and solving the fully vectorial monochromatic Maxwell's equations in 3D. The structures considered here are embedded in a larger computational domain surrounded by perfectly matched layers. Figure \ref{fig:oei} shows the electric field amplitude $|\bf E|$ for the different cases considered. To implement the OEi method, we determine the  matrix operator with the help of equation~\eqref{eq:matrix}. Here, we use angular spectral decomposition~\cite{Richards:1959p2607} of the incident light field corresponding to a numerical aperture of NA=0.8. 

In the first example (Fig. \ref{fig:oei} a-c), we use the intensity operator associated to the measure  $m^{(0)}$  to determine the largest transmission through a sub-wavelength aperture (diameter=200nm) in a thin layer of silver (thickness=200nm). The incident light field considered is linearly polarised and the transmission is determined across the output surface of the aperture. 

The electric amplitude $|\bf E|$ of most efficient transmission OEi is shown in Fig. \ref{fig:oei}a illustrating scattering of the aperture. The OEi on its own in Fig. \ref{fig:oei}b. The transmission enhancement factor, with respect to the tightest Bessel beam achievable for a numerical aperture of NA=0.8 (Fig. \ref{fig:oei}c), is 2.1 and 1.55 with respect to the Airy diffraction limited disk with the same numerical aperture. 

In the second example, we use the optical momentum operator associated with the quadratic measure $m^{(\mathbf{F\cdot u_z})}$ as defined in Table~\ref{tab}. The momentum OEi with the largest positive eigenvalue corresponds  to the field profile (Fig.  \ref{fig:oei} e-f) giving the largest optical force on the microparticle.  Figure  \ref{fig:oei}d shows the field amplitude $|\bf E|$ of this OEi on its own and scattering of the microparticle (Fig.  \ref{fig:oei}e). The optical force enhancement factor, with respect to the plane wave illumination (Fig. \ref{fig:oei}f), is of 49.3 and of 1.33 with respect to the Airy diffraction limited disk with the same numerical aperture.

\section{Discussion and Conclusion}

We have experimentally and theoretically demonstrated an approach based on optical eigenmodes that enables the minimization of the free space spot size of a beam. Using full vectorial simulations in 3D, we have shown how the OEi approach can be used to optimise light-matter scattering interactions in the case of transmission through sub-wavelength apertures and optical forces on micro-particles.  The generic nature of our approach means that it can be applied to other cases where the measure has a quadratic form and the propagation is linear.  In the present paper we have verified the rigor of the method by demonstrating the experimental spot size operator and intensity operator optimization using Laguerre-Gaussian and Bessel light modes using a dual SLM to implement the technique. Future work will aim to extend this method to optimise the size and contrast of optical dark vortices, the linear Raman scattering or the fluorescence of different samples, the optical trapping force, and the angular momentum transfer in optical manipulation.

\section*{Acknowledgements}
We thank the UK EPSRC for funding  through the Nanoscope Basic Technology grant. Mark Dennis is acknowledged for the introduction to super-oscillations. KD is a Royal Society-Wolfson Merit Award Holder. 

\section*{Annex: Dimensionality study}

In this appendix, we verify the convergence properties of our optical eigenmode method as a function of the dimensionality of the probing base used. Here, we consider a plane wave basis corresponding to the angular spectral decomposition of any incident field. The propagation is modelled using direct Fourier transform optics between the SLM plane and the target plane. Additionally the phase front of the incident beam is randomly changed to simulate the effect of high aberrations within the optical train. 
\begin{figure}[htb]
\centering\includegraphics[width=6.5cm]{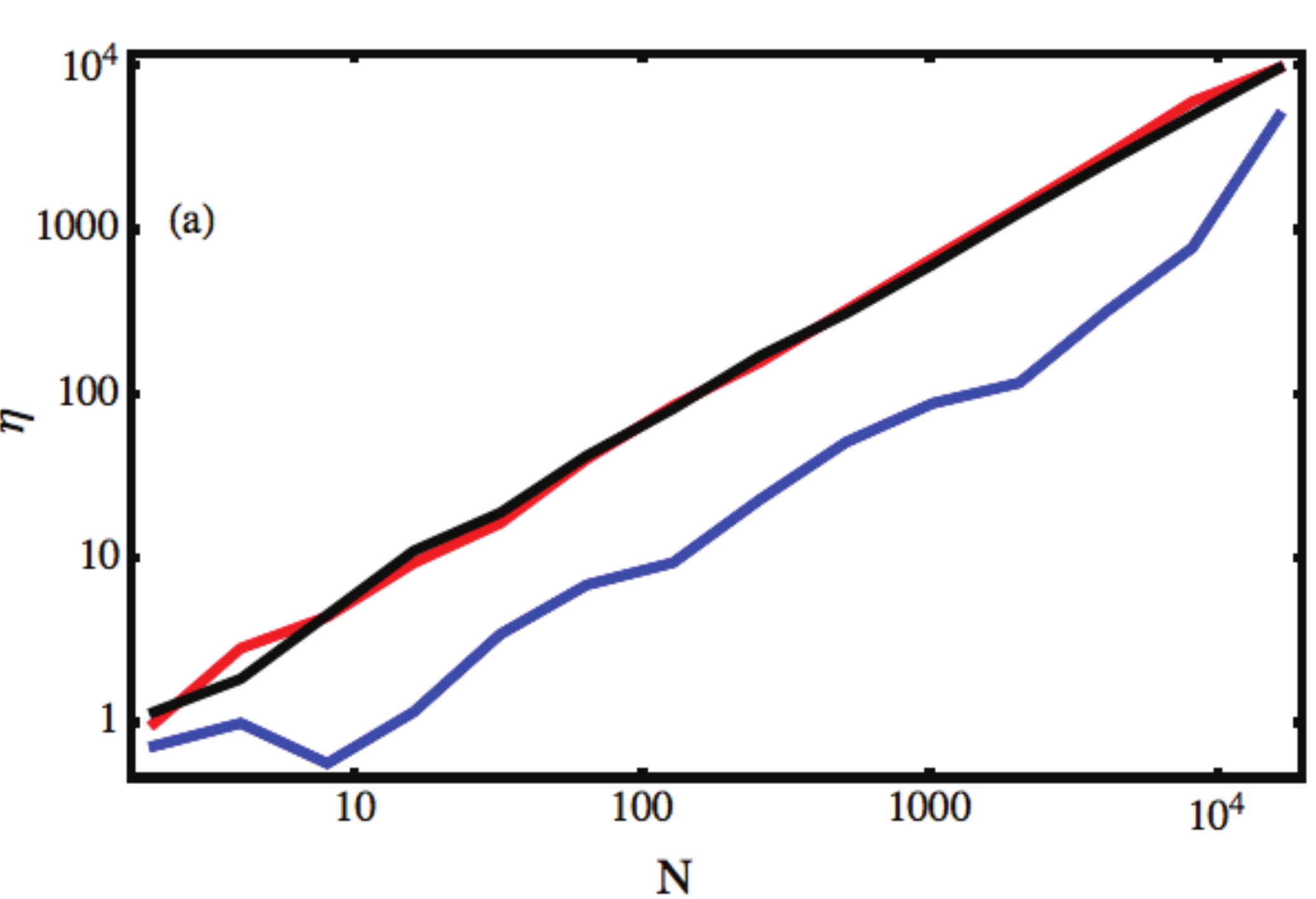}~~\includegraphics[width=6.5cm]{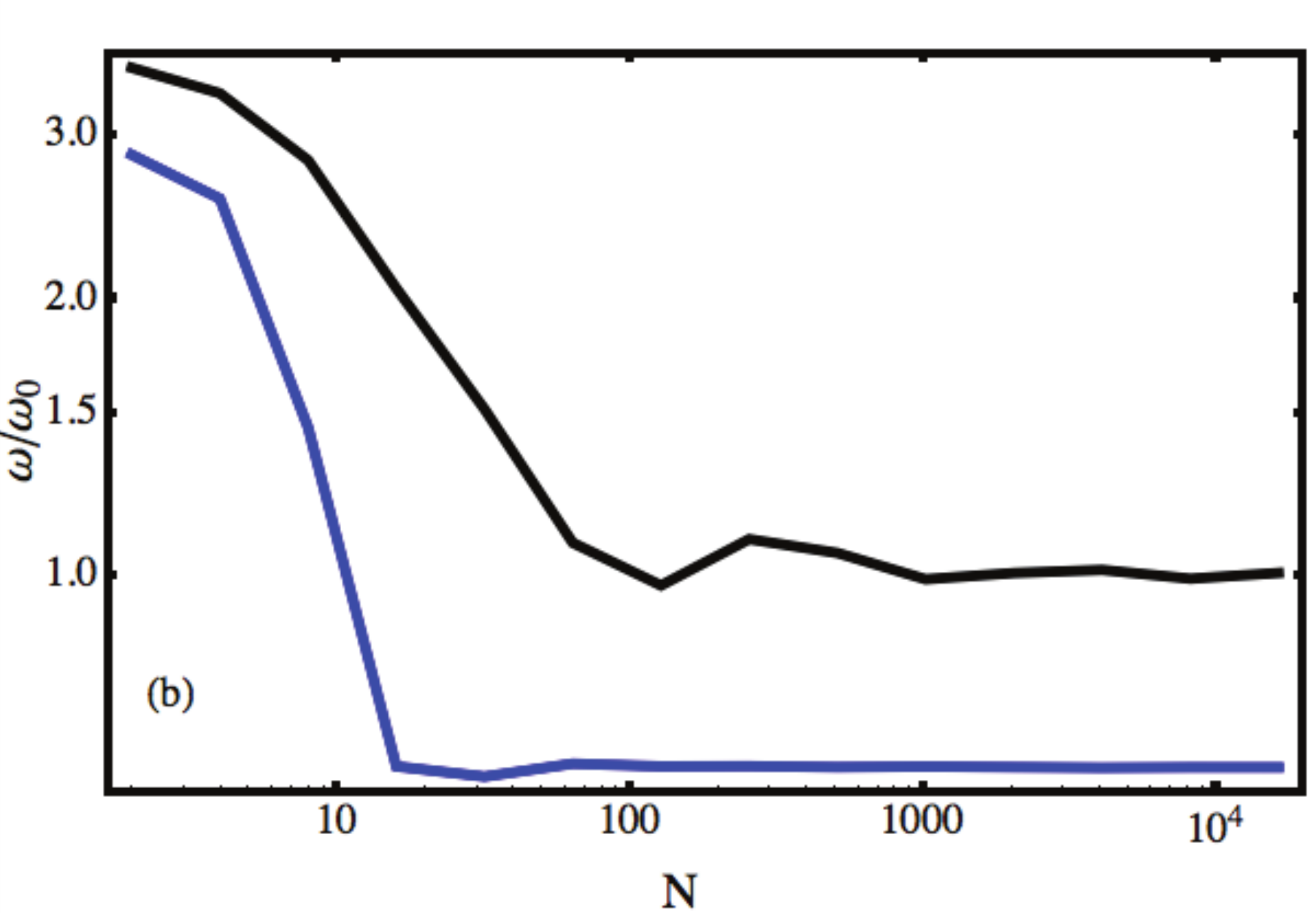}
\caption{\label{fig:conv}(a)  Comparison between the enhancement factor achieved using standard phase front correction techniques (in black \cite{Vellekoop:2008p9987,Cizmar:2010p9920}), the phase front correction from the intensity eigenmode (in red) and from the smallest spot size eigenmode (in blue). (b) Normalised beam spot size for the standard phase front correction techniques (in black \cite{Vellekoop:2008p9987,Cizmar:2010p9920}) and from the smallest spot size eigenmode (in blue). $\omega_0$ corresponds to the spot size of the Airy disk.}
\end{figure}
The dimensionality for the optical eigenmode method corresponds to the number of plane waves, $N$, taken into account, whilst retaining a fixed numerical aperture. The convergence behaviour is compared to standard phase front correction methods which can also be used to achieve focalised spots in the case of highly aberrated light fields. More precisely, these methods are based  on the variable partitioning of the SLM to create $N$ beamlets whose phases are individually changed such that all beamlets constructively interfere in the focal target point~\cite{Vellekoop:2008p9987,Cizmar:2010p9920}. This approach delivers a final correction phase mask that is able to correct for aberrations in the SLM incident light field additionally to the pre-correction of the propagation aberration between the SLM and the target focal plane~\cite{Cizmar:2010p9920}. This correction mask delivers an enhanced focal intensity $\eta$ shown in black in Fig.  \ref{fig:conv}a for both standard methods~\cite{Vellekoop:2008p9987,Cizmar:2010p9920}. For comparison purposes, we define a phase front correction mask using the phase part of the intensity operator eigenmode (red in Fig. \ref{fig:conv}a) which shows that, for all three methods, the enhancement factor scales linearly with the dimension $N$. 

Further, Fig \ref{fig:conv}b shows that the standard approaches cannot beat the diffraction limit, given by the Airy disk. In stark contrast, the smallest spot size eigenmode delivers sub-diffraction over almost the whole dimensional range considered with increasing efficiency (blue curve in Fig. \ref{fig:conv}a) for increasing dimensionality.

\end{document}